\documentclass[aps,prl,twocolumn,groupedaddress,showpacs,fleqn]{revtex4-1}
\usepackage{amsmath}
\usepackage{graphicx}
\usepackage{epstopdf}
\usepackage{epsfig}
\usepackage{amssymb}
\usepackage{bm}
\newcommand{\beq}{\begin{equation}}
\newcommand{\eeq}{\end{equation}}

\usepackage{color}

\begin{document}

\title{Singular Value Decomposition, Hessian Errors,  and Linear Algebra of Non-parametric Extraction of Partons from DIS}
 %for error analysis?
 
\author{Mehrdad  Goshtasbpour} 
%\email{Goshtasb@sbu.ac.ir}
\affiliation{Dept. of Physics, Shahid Beheshti University, G.C., Evin 19834, Tehran, Iran}
 
\begin{abstract}

 By singular value decomposition (SVD) of a numerically singular Hessian matrix and a numerically singular system of linear equations for the experimental data (accumulated in the respective ${\chi ^2}$ function) and constraints, least square solutions and their propagated errors for the non-parametric extraction of Partons from $F_2$ are obtained. 
 
   SVD and its physical application is phenomenologically described in the two cases.  Among the subjects covered  are: identification and properties of the boundary between the two subsets of ordered eigenvalues corresponding to range and null space, and the eigenvalue structure of the null space of the singular matrix, including a second boundary separating the smallest eigenvalues of essentially no information, in a particular case. The eigenvector-eigenvalue structure of "redundancy and smallness" of the errors of two pdf sets, in our simplified Hessian model, is described by a secondary manifestation of deeper null space, in the context of SVD.
\end{abstract}
\maketitle
\section{Introduction}
Singularity is widespread in nature and intriguing. %, as viewed by ,, 'is->,' ...mind to model nature.
  Linearization is a major tool of human mathematical mind. In many areas of science and engineering, in  particular, in physics, 
the enchanting theorem of linear algebra and the algorithm embodying it, the Singular Value Decomposition (SVD) of singular matrices is finding its applications. 
We encountered singularity in the way of extraction of parton distributions directly from the structure function $F_2$, in deep inelastic scattering (DIS), first in \cite{GS}. 

Based on one of the possibly typical examples, we are having an exposition of the physics of SVD, of how it diagnoses singularity in a linear system without ambiguity, and how it allows for its removal, bringing out desirable physical answer from its hiding place in the singular linear system. The example is a numerically singular Hessian matrix $(H)$ error analysis in the context of a numerically singular system of linear equations for the least square (LS)  estimates, of non-parametric extraction of parton distribution functions (pdfs),  from experimental data, $F_2$, accumulated in a ${\chi ^2}$, plus the equations of constraints, \cite{11084932v1}.
\subsection{Notation and definitions}
%\textbf:\\
$e_k={h_k}/ \sqrt{{\lambda}_k}$, defined under $h_k$;\\
$f=n_F+1\leq 6$: no. of pdf sets, ordered as 
in TABLE \ref{tab:2}; \\
$H (or A)$: numerically singular Hessian (or coefficient) matrix;\\
${h_k, k=1,..., R}$: set of orthonormal 
 eigenvector basis of the range of $H$, corresponding to the eigenvalues ${\lambda}_k, k=1,..., R$; 
$e_k={h_k}/ \sqrt{{\lambda}_k}$;\\
${\lambda}_k$, defined under $h_k$;\\
$m=f\times n-1\leq 65$: dimension of pdf-variables space;\\
 $n=11$: no. of $x-$bins of BCDMS \cite{BCDMS}, TABLE \ref{tab:2}, wherever not specified in the context;\\
$n_F \leq 5$: no. of flavors;\\
$Q^2=37.5$ $Gev^2$: arbitrarily chosen scale for LS estimated solutions;\\
$R $: dimension of range of $H (or A)$;\\
r-n border: boundary between the two subsets of ordered eigenvalues corresponding to range and null space of $H (or A)$;\\
$\sigma_i=\Delta{u_i}= u_i-u^0_i, i=1,...,m$: one standard deviation errors, (\ref{3-3});\\
$u_i, i=1,...,m$, $U=(u_1, ..., u_m)$: pdf-variables at $Q^2$,  $u_1=q_3(x_1)$, ..., $u_n=q_3(x_n)$, $u_{(n+1)}=q_8(x_1)$, ..., $u_m=g(x_n)$ in order of TABLE \ref{tab:2},
least square (LS) estimated at $U^0=(u^0_1, ..., u^0_m)$,  given in TABLE \ref{tab:2}; \\
$u^0_i$, with
$i=((j-1)\times{n}+k)$ before $q^{24}(x_{11})$ is dropped (due to lack of data), and $i=((j-1)\times{n}-1+k)$ afterwards, $j=1, ..., f$,  $k=1, ...,n$ except for $j=4$ ($q^{24}$): 
value of solution for kth
 element of the ${j}$th pdf set 
  at $(x_k, {Q}^2)$, given in TABLE \ref{tab:2}, alternatively defined under $u_i$; \\
 $z_k, k=1,..., R$: appropriately normalized, orthogonal directions in the range %of the 
 space of  $\Delta{u_i}$, defined in (\ref{3-1}) - (\ref{3-3}); \\
ZM VFN equations: zero mass variable flavor number equations of continuity constraints (for matching pdfs at $Q^2=m_b^2$ and $m_c^2$).
 \section{${\chi ^2}$}
Generally, for a set of $n$ analyzed experimental $F_2^{exp}$ data points with a given covariance matrix $V_{exp}$,
  \begin{equation}\label{1-1}
{\chi ^2} = \sum\limits_{i,j=1}^{n}{\Delta{{F_2}^{i}}{({V_{exp}^{-1}})^{ij}}\Delta{{F_2}^{j}}},
 \end{equation}
where $\Delta{{F_2}^{j}}={F_2^{j,exp}}-{F_2^{j,th}}$, and the theory (th) is LO PQCD here containing our pdfs directly as would be parameters, whose least square (LS) estimates minimizes ${\chi ^2}$. In (\ref{1-1}), obviously, constraint equations are left out of ${\chi ^2}$. 

There is a formal similarity between (\ref{1-1}) and a Taylor expansion of $\chi^2 $ about its LS minimum: 
 \begin{equation}\label{1-2}
\Delta{\chi ^2} = {\chi ^2}-{\chi ^2}_{min} \approx\sum\limits_{i,j=1}^m {\Delta{u_i}({{V_{u}^{-1}})_{ij}}\Delta{u_j}}.
 \end{equation}
  In (\ref{1-2}), $V_{u}$ is the covariance matrix of the unknown pdf-variables $u_i,  i=1, ..., m$, here replacing the "pseudoinverse" of the numerically singular Hessian matrix of second partial derivatives of $\chi ^2$ (\ref{2-5}); $\Delta{u_i}$ is the respective variation of $u_i$ about its least square (LS) estimate, $u^0_i$ - or the error.
  
   We may always compute the Hessian $H$, (\ref{2-5}), and use (\ref{1-2}); however, we are not often given a full experimental covariance matrix. It is diagonal, having independent measurements, and (\ref{1-1}) reduces to the following form that we use: 
  \begin{equation}\label{1-3}
{\chi ^2} ={\chi _p^2}+{\chi _d^2}=
 \sum\limits_{i,j=1}^{n_p}\frac{{(\Delta{F_2}^{p}_i)}^{2}}{{(\sigma^p_i)}^{2}}+
 \sum\limits_{i,j=1}^{n_d}\frac{{(\Delta{F_2}^{d}_i)}^{2}}{{(\sigma^d_i)}^{2}}.
 \end{equation}
 
 In our present analysis, in (\ref{1-3}), $n_{p,d}=153, 146$ are the total number
  of ${p,d}$ data left for analysis of BCDMS \cite{BCDMS} 
   after application of typical LO PQCD cuts for higher twists at the invariant mass squared, or squared boson-nucleon center-of mass energy $W^2=20 Gev^2$ \cite{MSTW}. 
   
 \section{Singularity and SVD of ZM VFN LO Hessian and other Matrices from ${\chi ^2}$ for ${F_2}^{p,d}$}  
 Singular Value Decomposition of a matrix decomposes it into three: the middle one contains the eigenvalues, the others contain the eigenvector bases of the subspaces of interest of the singular matrix for separating and managing the singularity, namely null space - on the right, and range - on the left \cite{Press}. 
The magic of SVD is to pinpoint numerically too small, ignorable, eigenvalues, corresponding to a ignorable set of eigenvectors of the numerical null space within domain. Remaining eigenvalues correspond to the remainder of domain which maps to the range, of the same dimension, in short, they correspond to range. Null space of $H$ is the subspace corresponding to the set of largest eigenvalues of its pseudoinverse, the covariance matrix $V$, which is side stepped or ultimately deleted, as the site of singularity. ${\chi ^2}$, carrying the physical information of both our theory and the experimental data ($F_2^{exp}$, justifiably assumed to be decisive in forming the numerical range), is decisive in determining the boundary between the two subsets of eigenvalues corresponding to range and null space, here called "r-n border", for all the singular matrices obtained via its derivatives. 

Operationally, i.e. in the process of trial motion up or down the scale of eigenvalues in the simulation program (determining R in (\ref{3-3})), there is a single physical criterion for uncovering the r-n border of $H$: appearance of well patterned set of errors for the LS estimate of every pdf set.  
For $H$, or other  singular matrices of the theory, r-n borderline acts as a "\textbf{physical ordering lens}", suddenly focusing the desirable physical properties of the solutions. We'll be using this concept as the \textbf{first} characterization or "qualitative, intuitive, physical definition" of the r-n border of a singular matrix.
The direction of motion from r-n border towards null space has a more immediate and dramatic effect on the mal-formation of the error groups compared to the other way. However, even the direction of motion towards the range has clear observable indications favoring optimal uncovering of the r-n border.
 
 Matrices of the theory are constructed from $\chi^2$, or ${\chi_i^2}, i=p,d$, (\ref{1-3}). For each, the most notable are two numerically singular matrices to begin with. Soon we will see that they may be made one and the same. One of them is the matrix of coefficients $A$ of the linear system of equations of 1st derivatives and the VFN constraints, to obtain the LS estimates of the pdf-variables, \cite{GS, 11084932v1}: 
   \begin{equation}\label{2-4}
A_{ij}{u_j}=b_i, \\ i=1,...,m+(2n-1); \\ j=1,...,m. 
 \end{equation}
 The other is the Hessian matrix of 2nd derivatives $H$:  
   \begin{equation}\label{2-5}
H_{ij}=\frac{1}{2}\frac{{\partial}^2 {\chi^2}}{{\partial {u_i}}{\partial {u_j}}}, i,j=1,...,m. 
 \end{equation}
  For BCDMS ${\chi^2}$, or ${\chi _p^2}$, (\ref{1-3}), $H$, (\ref{2-5}), is a square matrix on an $m=f\times n-1= 65$ dimensional pdf-variable space. For ${\chi_d ^2}$, $q^{3}$ is absent, thus $f=5$, and $m=54$. Subtraction of the number $1$ in $m$ is due to the fact that BCDMS data has no information on $q^{24}(x_{11}=.07)$, or on the quark $b(x_{11})$, as there is no data point with $Q^2\geq{m_b^2}$ at $x_{11}=.07$. (This statement is true in our simple "zero mass" VFN scheme.) Thus, $q^{24}(x_{11})$ is dropped from the set of variables and the set of constraint equations in (\ref{2-4}) and(\ref{2-5}). Dropping the VFN constraint equation with the missing $q^{24}(x_{11})$ leaves $(2n-1)$ equations of VFN constraint, at $m_c^2$ and $m_b^2$, thus resulting in an $[m+(2n-1)]\times m$ rectangular $A$ matrix.
  
  \begin{figure}[ht]
\begin{flushleft}
\includegraphics[height=.175\textheight, width=\columnwidth]{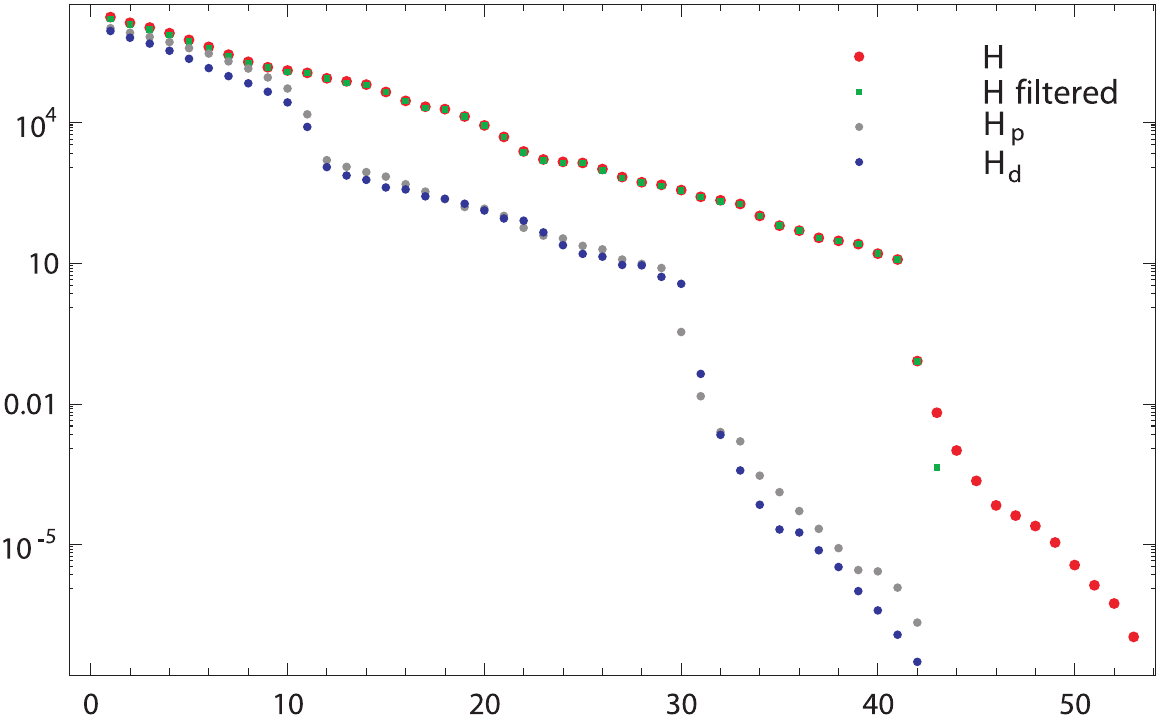}
\caption{Eigenvalues of Hessians}
\label{1}
\end{flushleft}
\end{figure}

\begin{figure}[ht]
\begin{flushleft}
\includegraphics[height=.175\textheight, width=\columnwidth]{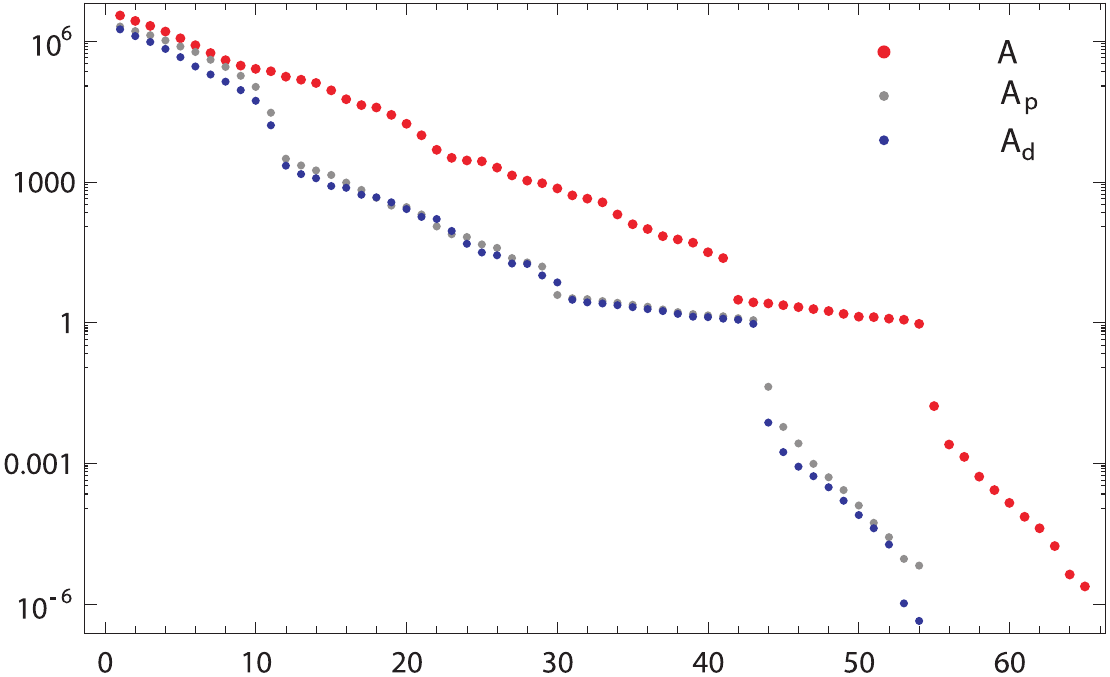}
\caption{Eigenvalues of all Coefficient Matrices A (\ref{2-4})}
\label{2}
\end{flushleft}
\end{figure}

For $H$ (of ${\chi^2}$), the r-n border, as a "physical ordering  lens", takes its place at $R=41$, corresponding to eigenvalues  of range ${\lambda}_k, k=1,..., R$ in the interval $[11.92, 1.77\times10^6]$, and a $24-$dimensional %deletable 
"null space", half (12) corresponding to relatively larger eigenvalues in the interval $(10^{-7}, .08]$ and the other half (12) to eigenva1ues numerically zero (within the accuracy of computation), figure \ref{1}.
There is considerable information in the 1st half of the null space, and very little in the second half, discussed  in details later. 

The \textbf{characteristic break} 
seen in figure \ref{1} at the r-n border of all the Hessians is a main lesson of  nature here, being translated to the act of a "physical ordering  lens", we had noted prior to seeing the curves. We use it as a \textbf{second}, and "quantitative definition of the r-n border".

In figure \ref{1}, a similar behavior of $H_p$ and $H_d$ can be seen, except for $R_p=29$ and $R_d=30$, reflecting having less information, loosely speaking, than $H$ with  $R=41$.  

Having a quadratic $\chi^2$ for $U$ everywhere in the $m=65$ dimensional pdf-variable space, twice the matrix of second derivatives, $H$, coincides with the data part of the matrix of coefficients $A$ of the linear system of equations, constructed from the 1st derivatives of $\chi^2$.

Thus, twice the eigenvalues of $H$ in figure \ref{1}, are the same as the eigenvalues of the data part of $A$, figure \ref{2}. The extra $13$ (or $14$ for $A_p$) eigenvalue tail, of the range of $A$, is due to the 21 constraint equations in $A$, not in $H$. 

\begin{figure}[ht]
\includegraphics[height=.175\textheight, width=\columnwidth]{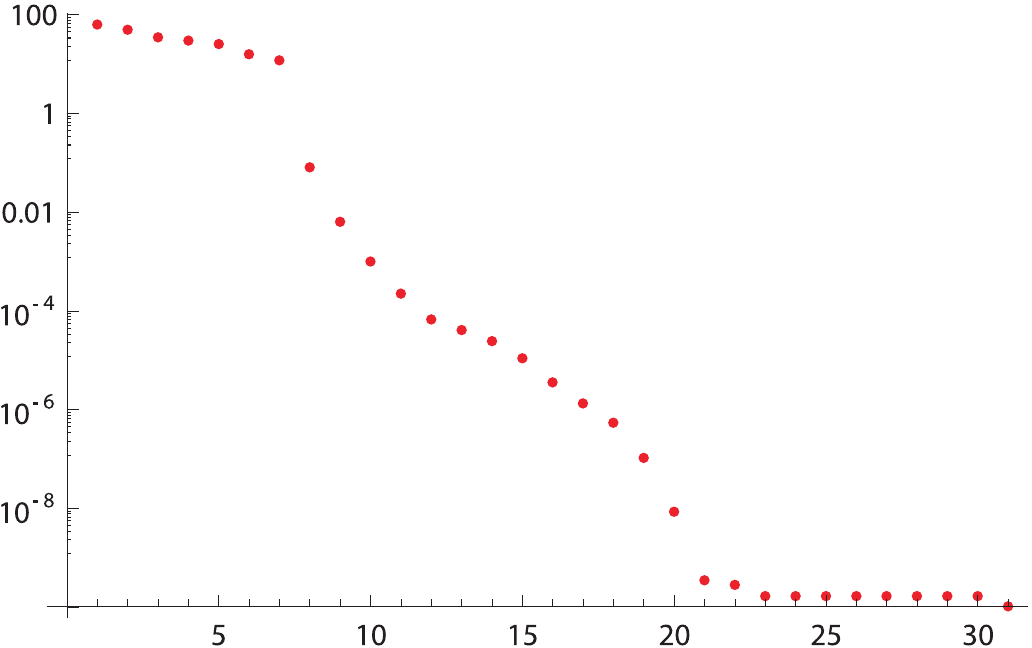}
\caption{31 smallest eigenvalues of $H$, null space, "r-n" and "redundancy" borders}
\label{3}
\end{figure}

A particular property of the r-n borderline is that its eigenvalues are of the order of $r=1$ to $10$ for all matrices of the model for all the subsets of the BCDMS data we encounter. It is interesting that the order of the eigenvalues in our range is similar to that of figure 2 of \cite{CTEQ1}, $(r, 10^6)$, in spite of our highly different contexts.

 The non-singular matrices of this data analysis arise from ${F_3}={F_2}^{p-n}$ for direct determination of the triplet pdf ${q^{3}(x_i, {Q}^2)}, i = 1,...,n.$\\  
 
 \section{Simple Linear Algebra and 
 Errors}
 Our simple linear algebra everywhere in the $m$ dimensional pdf-variables space, a quadratic $\chi ^2$ everywhere, exact expansion about its minimum (\ref{1-2}), constant Hessian, has the advantage that renders our method simpler than the methods of parametric determination of pdfs \cite{MSTW, CTEQ1}. Our choice of direct pdf-variables instead of their parameters, permitted through the linear algebraic matrix method of evolution - method of solution of DGLAP \cite{GS, 11084932v1} - allows a simple linear decomposition of $F_2$ in terms of $m$ pdf-variables - instead of their highly non-linear parameter dependence. The foundation of having a simple linear algebra everywhere can be seen to be:
 \begin{enumerate}
\item having $m$ direct pdf-variables, instead of intermediate parameters, 
\item having our matrix method for their evolution, 
\item having SVD for dealing with coefficient ($A$) and Hessian ($H$) matrices, and obtaining physically acceptable solutions. 
 \end{enumerate}

However, there is a limitation in our simple linear system which can be compared with the methods of parametric determination of pdfs for dealing with a single data set, BCDMS.
More specifically, in the error analysis of our $m$-dimensional singular Hessian, for the errors $\sigma_i, i=1,..., m$, it is the $R$- dimensional range which contains the main information. Thus there is some lack of information on the remaining $(m-R)$ errors! We can  
extract some of the information of the null space by a procedure in the lines that follow.

In line with \cite{CTEQ1} and  \cite{MRST2007} modified by SVD, Let ${h_k, k=1,..., R}$ be the set of 
orthonormal 
 eigenvector basis of the range of $H$, corresponding to the eigenvalues ${\lambda}_k, k=1,..., R$, each having m elements. $m=65$ and $R=41$ for H of figure \ref{1}. Scale the set as: 
 $e_k={h_k}/
 \sqrt{{\lambda}_k}, k=1,..., R$. Expand  
 the variation $\sigma_i$ about the LS estimate, $u^0_i$, in terms of the scaled basis,
 \begin{equation}\label{3-1}
\sigma_i=\Delta{u_i}={u_i}-{u_i}^0=\sum\limits_{k=1}^{R}{e_{ik}z_k}, i=1,..., m,
  \end{equation}
where $e_{ik}$ denotes the $i$th component of the vector $e_k$. Thus (\ref{1-2}) implies
 \begin{equation}\label{3-2}
\begin{array}{l}
\Delta{\chi ^2} ={\chi ^2}-{\chi_{min} ^2}={\chi ^2}(U)-{\chi ^2}(U^0)=% \approx\\
\\
\sum\limits_{i,j=1}^m {\Delta{u_i}({{V_{u}^{-1}})_{ij}}\Delta{u_j}}= \sum\limits_{k,l=1}^R\sum\limits_{i,j=1}^m{z_ke^t_{ik}H_{ij}e_{jl}z_l} \\
=\sum\limits_{k,l=1}^R z_k z_l({{\lambda}_l}/
 \sqrt{{\lambda}_k{\lambda}_l})\sum\limits_{i=1}^m h^t_{ik}h_{il}=\sum\limits_{k=1}^R{z_k^2}.
\end{array}
 \end{equation} 
 $z_i$ are $R$ appropriately normalized linear combination of $\Delta{u_i}$, defining orthogonal directions in 
  the space of deviations of pdf-variables. By (\ref{3-2}), a surface of constant ${\chi ^2}$ is a hyper-sphere of given radius in $z$-space.
 The error of the quantity $u_i$ in $z$-space is, equation (7) of \cite{MRST2007},
 
\begin{equation}\label{3-3}
\sigma_i=\sqrt{\Delta{\chi ^2}.{\sum\limits_{k=1}^{R}({\frac{{\partial}{u_i}}{{\partial {z_k}}}})^2}}=\sqrt{\sum\limits_{k=1}^{R}{e_{ik}^2}}, i=1, ..., m.
  \end{equation}     
% also eqs: (15), (30) \cite{CTEQ1}
where $\Delta{\chi ^2} =1$ was used, corresponding to one standard deviation errors, or $68$ percent confidence level.    

There are only $R$ degrees of freedom in terms of $z_i$. However, we sought $m$ solutions of the errors in (\ref{3-3}). Thus there is a complication for $(m-R)=24$ of the errors. This appears as $24$ small errors, $22$ of which are in $2$ redundant error sets, discussed in the next section.
 
Having singularity and using SVD, the major difference with references \cite{CTEQ1} and  \cite{MRST2007} is limitation by $R$, for r-n border, in (\ref{3-1}) through (\ref{3-3}), dictated by the criteria of the previous section which boils down to the numerical observation that the errors become highly unacceptable, as soon as any eigenvector of the null space is added. 

Equation  (\ref{3-3}) constructs errors which are grouped, in \textbf{Well formed}, i.e, dominantly increasing, sets for each of the $f$ pdf sets. It may be understood as follows. For the $l$th element of the $J$th pdf set, in the prescribed order of TABLES \ref{tab:1} and \ref{tab:2},  the error $\sigma_i$, is defined in (\ref{3-3}), with  $i=((j-1)\times{n}+l)$, before $q^{24}(x_{11})$ is dropped, and $i=((j-1)\times{n}-1+l)$ afterwards. Each error $\sigma_i$, being the $i$th component of the scaled eigenvectors $e_k$ as summed in (\ref{3-3}), is observed to be dominated by one, or more $e_k$, bunched together within the range, at a distance, correlated with $(n-l)$, roughly $\geq{(n-l)}$, from the r-n border. It is so that, for each $j$, increase of $l=1,..., n$ ($n\rightarrow(n-1)$ for $q^{24}$) on the average decreases the distance of the dominant $e_k$ bunch from the r-n border, creating the dominantly increasing trend of the errors, within the jth set, because of
 scaling of $e_k$. This structure is imparted to the errors via $\chi^2 $ through the Hessian and eventually, through the mentioned structure of the index $i$ in (\ref{3-3}).

Note that if $1/2$ of the coefficients of the $(2n-1)$ linear equations of VFN constraints were added to $H$, making it  
$[m+(2n-1)]\times m=86\times 65$ dimensional, $H$ would have exactly $1/2$ of the whole spectrum of $A$ of figure \ref{2}.
 However, %that has no physical meaning. 
  such a simple addition of coefficients of constraints to $H$  is not allowed, as it is the matrix of second derivatives of $\chi^2$. From $H$ meaningful (\textbf{well-formed}) errors are calculated at the answer $U^0=(u^0_1, ..., u^0_m)$ at the chosen fixed $Q^2$ (here $37.5 $ $Gev^2$), having the effect of continuity of $U^0$ at matching points due to the constraints, but no continuity of their variations, the errors, which becomes an assumption in the determination of errors that follows. 

Addition of the coefficients of constraint equations to $A$ follows naturally. It leads to spread of the range at the r-n border as may be seen in figure \ref{2} by some extra eigenvalues, all of the order of $1$. Physically meaningful solutions of $A$ are obtained only through such addition, and then, as expected from figure \ref{2}, if the r-n border is placed at $R=54$ and $R_d=R_p=43$, so that the range includes the constraints. Deleting from $A$ the less critical  n constraints  at $Q^2=m_c^2$, as explicitly shown in \cite{11084932v1}, leads to non-physical answers.

 For $A$, (constraints for) continuity of pdfs are essential to get any physical results. It is not so for $H$, which allows doing without constraints, and having  a simplifying assumption. However, there are problems, some faced in the next section, very likely beginning in this simplification.
  
 \section{Redundancy and Smallness of Errors of two pdf sets}
 
 Using the Hessian (\ref{2-5}), there are problems of redundancy and having too-small error bars, associated with redundancy.
  Within our method, redundancy has a straight forward definition, referring to sameness of the quantity of error, i.e. (\ref{3-5}). 
   
   When the two separate SVD for matrices $A$ and $H$ are done, we are left with physically desirable LS estimate of the pdfs, for $d$ and $p+d$ data% ( for $p$ data set alone, there is a problem)
.  And there are \textbf{Well formed} sets of errors for each of the $f=5$ or $6$ groups of pdfs, for $d$ and $p+d$ data respectively, except for a "group redundancy" in the following sets: 
\begin{equation}\label{3-5}
 \Delta{q^{15}(x_i, {Q}^2)}= \Delta{q^{8}(x_i, {Q}^2)}, i=1,...,n, 
 \end{equation}
 as seen in TABLE \ref{tab:1}. The $2n=22$ redundant error bars, (\ref{3-5}), like the others, are well-formed, and acceptable except for being too small to be one standard deviation, for all of the different analysis, $p,d$, and $p+d$. 
 
 Indeed, details of redundancy happen to be:  
\begin{equation}\label{3-6}
 e_{(i+{p}\times{n}),k}=e_{(i+(p+1)\times{n}),k}, i=1,...,n, k=1,...,R;
 \end{equation}
 where %ordering of ... corresponds to that in "Notation ..." or in Tables I and II, and
  $p=1$ or $0$, depending on whether proton data is included or not. Equations (\ref{3-6}) and (\ref{3-3}), not only imply (\ref{3-5}), but also         bring algebraic rigor to the term "information" and  its shortage  in terms of the properties of eigenvectors of $H$.
 
   Redundancy or duplication of errors, lack of information, being at the bottom of the null space are associated phenomena.
  As other aspects of the null space in our model, it may be eliminated %\footnote{In the different context of global parametric data analysis, e.g. \cite{MSTW}, a similar vocabulary, "elimination of redundancy", is used}.
    \begin{table}
\begin{tabular}{|l|c|c|c|c|c|c|}
\hline
$x_i$&$\Delta{q^3_{p+d}}$& $\Delta{q^{8,15}_{d}}$&$\Delta{q^{8,15}_{p+d}}$&$\Delta{q^{24}_{p+d}}$
&$\Delta{q^s_{p+d}}$&$\Delta{g_{p+d}}$\\ \hline \hline
0.75 &0.00280 &0.00334 &0.00234 &0.00245 &0.00290 &0.0844 \\
0.65 &0.00301 &0.00437 &0.00379‎ &0.0205 &0.00511 &0.135 \\
0.55 &0.00344 &0.00764 &0.00742‎ &0.0449 &0.0101 &0.122 \\
0.45 &0.00388 &0.00785 &0.00590 &0.0341 &0.00907 &0.112 \\
0.35 &0.00447 &0.00863 &0.00593‎ &0.0349 &0.0105 &0.105 \\
0.275 &0.00648 &0.0109 &0.00728‎ &0.0473 &0.0154 &0.146 \\
0.225 &0.00712 &0.0115 &0.00759‎ &0.0519 &0.0176 &0.188 \\
0.18 &0.00853 &0.0117 &0.00816 &0.0561 &0.0198 &0.195 \\
0.14 &0.0106 &0.0133 &0.00929‎ &0.0628 &0.0225 &0.206 \\
0.1 &0.0134 &0.0190 &0.0130‎ &0.0789 &0.0238 &0.173 \\
0.07 &0.0298 &0.0135 &0.0102 & &0.0433 &0.330 \\
\hline
 \end{tabular}
\caption{One standard deviation Hessian errors for proton plus deuteron ($p+d$) data analysis, including redundancy and smallness of the errors of both $p+d$ and $d$ data ($\Delta{q^8}=\Delta{q^{15}}$ columns).}
\label{tab:1}
\end{table}
\begin{table}
\begin{tabular}{|l|c|c|c|c|c|c|}
\hline
$x_i$&${q^3_{p+d}}$&${q^{8}_{p+d}}$&${q^{15}_{p+d}}$
&${q^{24}_{p+d}}$&${q^s_{p+d}}$&${g_{p+d}}$\\ \hline \hline
0.75 &0.0138 &0.0625 &0.0326 &0.0171 &0.0161 &-0.127 \\
0.65 &0.0532 &0.0781 &0.0625‎ &0.0456 &0.0746 &0.274 \\
0.55 &0.117 &0.157 &0.135‎ &0.173 &0.172 &0.114 \\
0.45 &0.189 &0.294 &0.299 &0.346 &0.359 &0.376 \\
0.35 &0.272 &0.492 &0.514 &0.691 &0.588 &0.219 \\
0.275 &0.313 &0.644 &0.762‎ &0.861 &0.813 &0.137 \\
0.225 &0.326 &0.761 &0.914 &1.05 &0.955 &0.166 \\
0.18 &0.310 &0.880 &1.06 &1.26 &1.08 &0.0142 \\
0.14 &0.273 &0.970 &1.19‎ &1.33 &1.24 &0.478 \\
0.1 &0.230 &1.04 &1.27‎ &1.42 &1.41 &0.894 \\
0.07 &0.169 &0.744 &1.38 & &1.57 &1.01 \\
\hline
 \end{tabular}
\caption{ Least square estimated central values of the pdf solutions of equations (\ref{2-4}) for proton plus deuteron ($p+d$) BCDMS data analysis at $Q_0^2=37.5 Gev^2$, obtained at the $x$-points of the data.}
\label{tab:2}
\end{table}

  Exactly all of the eigenvectors of null  space were forced out of counting in the sum of (\ref{3-3}), to keep the formation of error sets well and acceptable. However, $R$ eigenvectors of range, each have $m$ components,  associated with the $m$ ordered eigenvalues, due to the symmetry of the ${m}\times{m}$ matrix of eigenvectors. Thus, the relation of the last $(m-R)$ components to the null space is to be taken into account now, leading to a completion of deletion of a bottom subspace of the null space. 

A \textbf{key} observation, on the behavior of null space in the physical application of SVD to the Hessian, is that elimination of the two sets of redundant and smallest error bars completely eliminates the deepest part of the numerical null space, corresponding to $2n$ smallest eigenvalues.  Elimination of %(an arbitrary) 
 a set of $n$ redundancies eliminates $n$ smallest eigenvalues. 
  Thus,
along with redundancy, $n$ dimensions are to be totally eliminated. However, the smallness of the remaining $n$ error bars is essentially untouched, whose elimination completely drops the next deepest $n$  dimensions of the null space of $H$. 

Indeed, generally, a mapping of error sets into the ordered eigenvalue spectrum of the Hessian may be obtained by an ordered dropping, beginning with one set at a time. It takes the following order for ${F_2}^{d}$ analysis: $q^s, q^{24}, g, q^{8, 15}$ (with an exception of 2 error points coming from above to below the $g$ set, right below the r-n border).  For ${F_2}^{p,d}$ analysis, bulk of the extra $q^3$ set (10 out of n=11) goes between $q^s$ and $q^{24}$, in the above ordering.

The only by-product of dropping a set, is a slight extension or stretching of the eigenvalues towards the void the drop creates, which dies off with distance from the void! 
In figure \ref{1}, omission of the smallest $2n$ eigenvalues, after elimination of $2n$ redundant and small error bars of total $H$ can be observed in $H_{filtered}$. 
It results is a slight shrinkage of the eigenvalues (towards the created void), becoming pronounced for the smallest remaining eigenvalue(s), best observable for the last, or  $(m-2n)th=43rd$, eigenvalue of $H_{filtered}$. 

\begin{figure}
\includegraphics[height=1.01\textheight, width=\columnwidth]{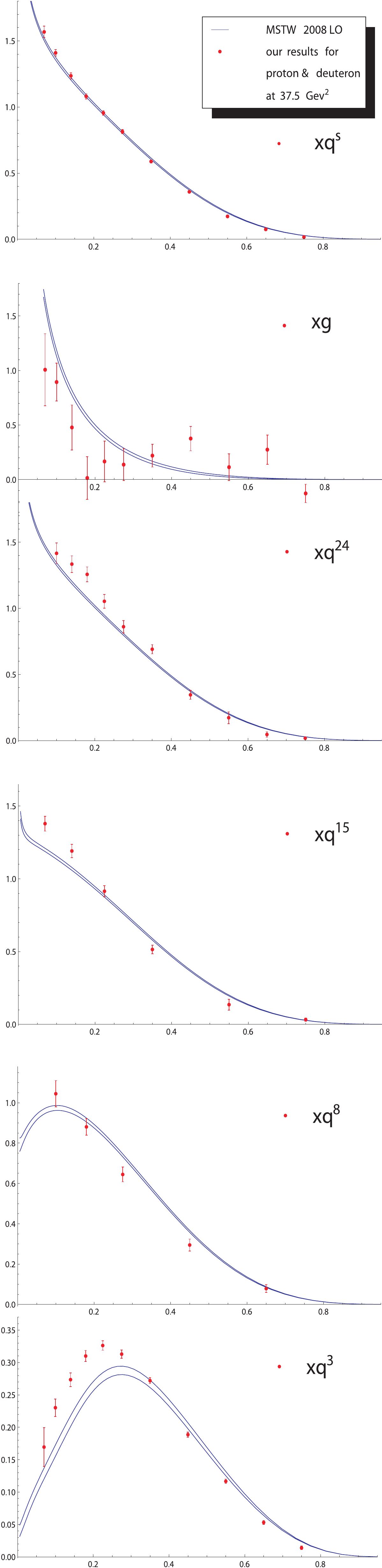}

\end{figure}

\begin{figure}
\includegraphics[height=.807\textheight, width=\columnwidth]{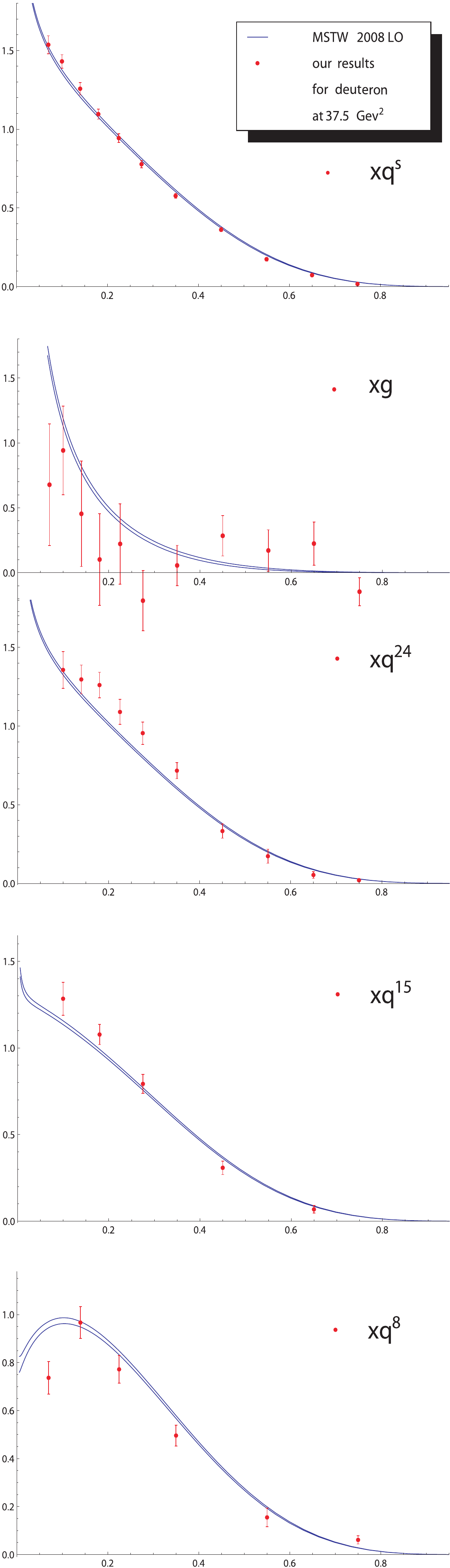}
\caption{Points are ZM VFN LO parton momentum distributions at $Q_0^2=37.5 Gev^2$, obtained at the $x$-points of the data. They have normal one standard deviation errors, except for $q^{8}$ and $q^{15}$, which arbitrarily have every other point missing (left and right show alternating $x$-points), to eliminate the redundancy of errors, and have the remaining error bars enlarged by a factor of $5$, to remedy the smallness. Continuous curves are the corresponding GM global MSTW LO distributions with error margin. 
On the left, there are six sets (five quark flavors) obtained from $299$ data points of BCDMS ${F_2}^{p,d}$ set.  On the right, there are five sets obtained from $146$ data points of BCDMS ${F_2}^{d}$ subset.
}% for JHEP NEED a division  of page into two columns for two figures left and right: E.O.
\label{4}
\end{figure}

Figure \ref{3} shows a continuation of figure \ref{1}, for the eigenvalues of $H$, obtained by raising the sensitivity of calculation. In it, the drop after r-n border is followed by a second drop before $n$ approximately zero redundant eigenvalues, we may
 call "redundancy border", beyond which there is essentially no information. 

Beyond redundancy and its associated smallness of errors, belonging to the bottom of the null space, there are other relatively small, nevertheless, justifiable error bars observable in TABLE \ref{tab:1} and figure \ref{4}. The smallest such error bars are $\Delta{q^{s}_{p+d}}(x_i, Q^2), 
i=1,2$ and $\Delta{q^{24}_{p+d}}(x_1, Q^2)$ (of which two are associated with the remaining highest two eigenvalues of null space, just below the r-n border), which, excluding inexact gluon, happen to be the largest relative errors, having the smallest LS central values; and the whole set $\Delta{q^{s}_{p+d}}(x_i, Q^2), i=1,...,n$, which is acceptable on the grounds that $SU(5)$ singlet is apparently the most exact parton distribution that our phenomenology (assumptions) gets out of this data, given the standard yard stick of LO MSTW. %why, given the $b$ problem of \cite{I}?
  
\section{Results and Discussion}
    One may remedy the relevant information limitation of the small error sets via a single \textbf{enlargement} ("tolerance" %\footnote{An alternative term used in global parametric data analysis, e.g. in \cite{MSTW}}
) factor $T$, 
 for each $p+d$ or $d$ data analysis, mainly as each $n$-element set of errors is \textbf{well formed}. 
Putting a $T$ before a too small set $\sigma_i, i\geq{R+3}$, is equivalent to having a $T$ before the $i$th component of the eigenvectors of $H$, (\ref{3-3}).
 
Figure \ref{4} left, as well as the TABLES, show six sets of pdfs, obtained from the analysis of $299$ data points of BCDMS  ${F_2}^{P,d}$ set. The graphs have normal one standard deviation errors, except for $q^{8}$ and $q^{15}$, with information lack in their errors, which arbitrarily have every other point missing, to eliminate the redundancy, and have the remaining error bars enlarged by a factor $T=5$, to remedy the smallness.  Figure \ref{4} right shows similar graphs resulting from the analysis of $146$ data points of BCDMS ${F_2}^{d}$ subset, where the sparse points of its bottom two graphs are opposite or complementary with  respect to those of figure \ref{4} left! Graphs include comparison with MSTW \cite{MSTW} pdfs with error margin.

Mismatch of the central values of our results 
with MSTW is due most likely to the non-asymptotic mismatch of VFN schemes, our ZM and MSTW's GM. Possibly, higher order interpolation of the type of HOPPET \cite{HOPPET} or QCDNUM \cite{Botje}, has an improving effect on our 1st order interpolation, as well. The effect of these factors on our pure finite evolution is discussed in further details in the second version of \cite{I}, to be submitted concurrently.
Beyond these considerations, the problem of computation of our errors includes that of having excluded VFN continuity of errors altogether. 

As figures \ref{4} left and right, as well as columns $3$ and $4$ of TABLE \ref{tab:1}, indicate, larger errors of the analysis of $146$ data points of  ${F_2}^{d}$ subset become, naturally, smaller as the pool of information is enlarged via addition of data to $299$ points of  ${F_2}^{p,d}$ set. Indeed, generally, for all of the errors, $\forall{i}$, ${1.0}\leq{\Delta{u^d_i}/\Delta{u^{p+d}_i}}\leq{1.8}$, with average, $\langle{\Delta{u^d_i}/\Delta{u^{p+d}_i}}\rangle =1.4$. Extension of this reasoning can apply to why the errors of the MSTW pdfs are well smaller.

Addition of equations of VFN constraints to the linear system (\ref{2-4}) results in calculation of LS estimates for the central values of the pdf-variables which do not minimize $\chi ^2$ of (\ref{1-3}) absolutely. Minimization is only on the hypersurface defined by the $2n-1=21$ equations of VFN constraints. ($\chi ^2(U_0)/d.o.f. =\chi ^2(U_0)/m\approx 2/3$.)

\section{Summary and Applications of SVD}
 We introduced a flexible linear algebraic direct method of determination of pdfs at the $x$-bins of the data,
 which has the advantage of simplicity.  
For our method, SVD is essential.
The basics of SVD in terms of the properties of numerical null space and its 
border with range are abstracted and expounded.
Our different cases of physical applications of SVD can be said to be essentially purging the null space in different ways, the most important being as the sight of singularity. 

First, we begin with our criteria of determining the r-n border to get the physical answers, out of their hiding place due to singularity, for both matrices $A$ and $H$. For $A$, eigenvalue consideration is enough, and determination of $R$ brings us physical solution of the linear system of equations with coefficient matrix $A$. For $H$, eigenvectors are brought in (\ref{3-3}). Those  of the null space are purged out of calculations to arrive at physical answers. However, the $R$ eigenvectors of range, each have $m$ components,  associated with the $m$ ordered eigenvalues, due to the symmetry of the ${m}\times{m}$ matrix of eigenvectors. Thus, the relation of the last $(m-R)$ components to the null space is analyzed. The last $n$ components are identified with the phenomenon of redundancy and with eigenvalues below redundancy border at the bottom of the null space, corresponding to the last $11$ points of figure \ref{3}. Not showing redundant error bars in our resulting graphs is taken to mean a deleting of the last $n$ components.

 SVD cut is not used to delete the remaining \textbf{well formed}, but too small error bar set, associated with redundancy. We try to keep whatever information which may be saved. There is an attempt to remedy the lack of information by a single degree of freedom corresponding to an enlargement factor T, equivalent to putting an arbitrary factor T before a set of $n$ deleted eigenvectors of the null space.
%, which is a freedom that is not shown in the graphs of eigenvalues of $H$.

%  Our simplifying assumption of neglecting to impose continuity of general VFN schemes on errors (and pdfs 1st derivatives) is probable cause of not having the other degrees of freedom, at least 12, needed for the complete set of errors.

 Other experimental and phenomenological linear numerical modeling may run into singular matrices.                     How the spectrum can lend itself to interpretation depends on use value in the context.  E.g. in such a linear modeling as signature of identification of molecules, a comparison of two singular spectrums with differences only in the bottom of the null space, of the type we have below the redundancy border in figure \ref{3}, is a signal that two molecules can replace each other.

"\textbf{Acknowledgements}:  

   I would like to thank my colleague M.S. Movahed for enlightening discussions, and note with pleasure the impulse  Ali Sadeghi provided for submission of this two year old work.

\end{document}